\documentclass[conference]{IEEEtran}
\IEEEoverridecommandlockouts
\usepackage{cite}
\usepackage{amsmath,amssymb,amsfonts}
\usepackage{algorithmic}
\usepackage{graphicx}
\usepackage{textcomp}
\usepackage{xcolor}
\usepackage{siunitx}  
\usepackage{listings}
\usepackage{bm}
\usepackage{subcaption}
\usepackage{tikz-cd}
\usepackage{tikz}
\usepackage{geometry}
\usepackage{setspace}

\def\BibTeX{{\rm B\kern-.05em{\sc i\kern-.025em b}\kern-.08em
    T\kern-.1667em\lower.7ex\hbox{E}\kern-.125emX}}
\begin{document}
\columnsep 0.205in

\title{Semantic Communication via Rate Distortion Perception Bottleneck}
\author{
\IEEEauthorblockN{Zihe Zhao, Chunyue Wang}
\IEEEauthorblockA{\small School of Communications Engineering, Jilin University, Changchun, China \\
Email:{ zhaozh2021@mails.jlu.edu.cn}, {chunyue@jlu.edu.cn}}
}

\maketitle

\begin{abstract}
With the advancement of Artificial Intelligence (AI) technology, next-generation wireless communication network is facing unprecedented challenge. Semantic communication has become a novel solution to address such challenges, with enhancing the efficiency of bandwidth utilization by transmitting meaningful information and filtering out superfluous data. Unfortunately, recent studies have shown that classical Shannon information theory primarily focuses on the bit-level distortion, which cannot adequately address the perceptual quality issues of data reconstruction at the receiver end. In this work, we consider the impact of semantic-level distortion on semantic communication. We develop an image inference network based on the Information Bottleneck (IB) framework and concurrently establish an image reconstruction network. This network is designed to achieve joint optimization of perception and bit-level distortion, as well as image inference, associated with compressing information. To maintain consistency with the principles of IB for handling high-dimensional data, we employ variational approximation methods to simplify the optimization problem. Finally, we confirm the existence of the rate distortion perception tradeoff within IB framework through experimental analysis conducted on the MNIST dataset.

\end{abstract}

\begin{IEEEkeywords}
semantic communication, rate-distortion-perception, information bottleneck.
\end{IEEEkeywords}

\section{Introduction}
During the exponential growth of Artificial Intelligence (AI), existing communication technologies have been significantly enhanced\cite{8054694,8642915}, and has greatly accelerated developments in the field of telecommunications\cite{8808168}. However, this progress also presents unprecedented challenges for current wireless networks. Applications such as Virtual/Augmented Reality (VR/AR)\cite{Inkarbekov_2023,doi:10.1080/00207543.2020.1859636}, autonomous driving\cite{atakishiyev2023explainable}, and medical diagnostics\cite{9279211}, have led to a relentless pursuit to enhance wireless communication bandwidth. This has spurred a significant amount of research aimed at enhancing wireless communication bandwidth. Numerous developments in wireless technologies are predicated on concepts originating from Claude Shannon’s seminal work on information theory, focuses on accurately delivering a set of symbols \cite{6773024}. With the aid of advanced encoding, decoding, and modulation techniques, modern communication systems are rapidly approaching the physical capacity limit delineated by Shannon’s theory. However, the present capacity limit still fails to meet the continuously growing demand for wireless communication bandwidth. Therefore, semantic communication, capable of surpassing the constraints of Shannon’s capacity limits, is considered a promising solution to address this challenge. Consequently, semantic communication has gained widespread attention from researchers and is regarded as a fundamental challenge for the sixth generation (6G) of communication \cite{9446676,9770094,10233481,9937052}. 

Approximately 75 years ago, in the article\cite{weaver1953recent}, Weaver categorized communication into three levels: syntactic communication, semantic communication, and pragmatic communication. He highlighted the main difference between the semantic communication problem and Shannon’s engineering problem, which is that the semantic communication paradigm provides a receiver with meaningful information extracted from the source. And lately the authors in \cite{Guler2016TheSC} introduced a framework for semantic communication aimed at reducing semantic discrepancies, which is grounded in the semantic theory of logical probability. Recent research in the field of semantic communication has shifted towards semantic entropy. In \cite{chattopadhyay2021quantifying}, the authors have quantified semantic entropy and the complexity involved in semantic compression. Through this way, researchers have glimpsed the potential to enhance system capacity, transcending the confines of Shannon’s capacity limits. Then in order to better realize the semantic communication, researchers frequently employ task-oriented communication, which is designed to extract and transmit information pertinent to the task at hand. Meanwhile, Joint Source-Channel Coding (JSCC) is frequently utilized for feature extraction and encoding in semantic/task-oriented communication. There has been considerable research into JSCC based on learning, which is a technique that utilizes deep neural networks (DNNs) to discern and extract relevant data, subsequently mapping the task-specific information onto a continuous channel input\cite{8461983,xie2024deep}.
\begin{figure*}[t]
\centering
\includegraphics[width=18cm]{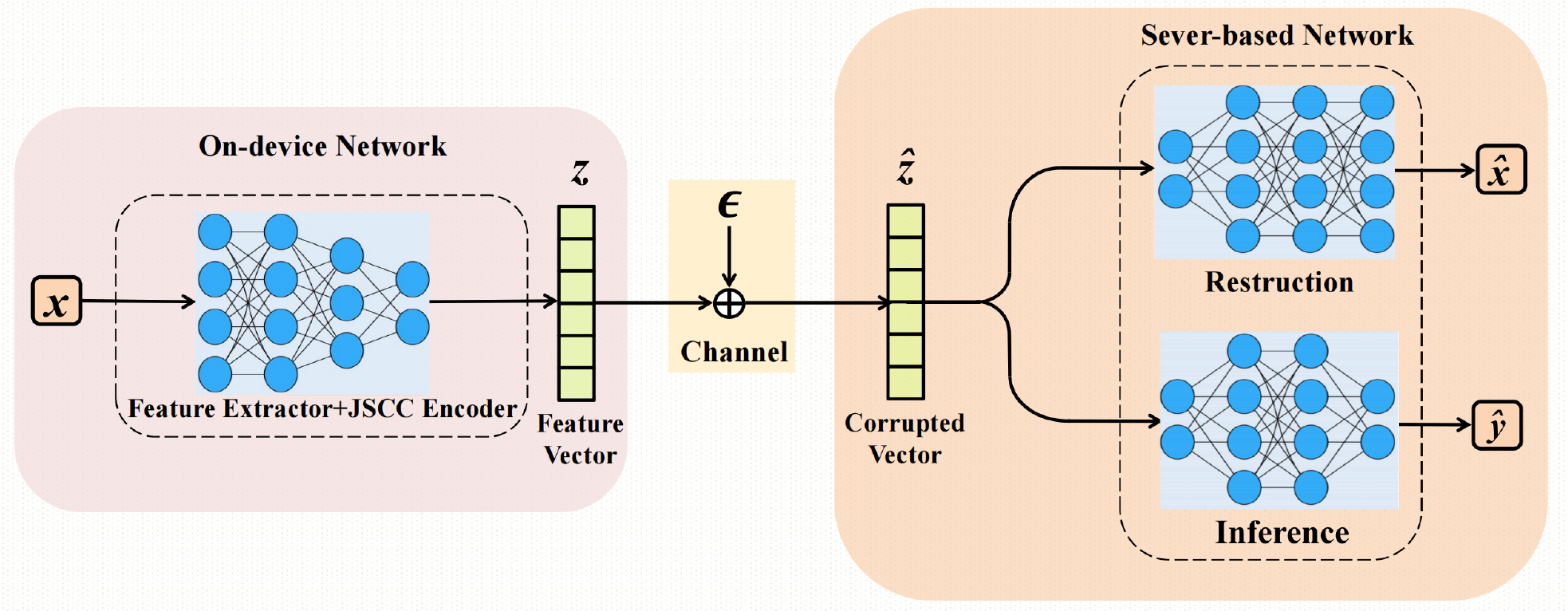}
\caption{ Semantic Communication model}
\label{ Semantic Communication model}
\end{figure*}

Nevertheless, bit-level distortion metrics such as Mean Absolute Error (MAE) and Peak Signal-to-Noise Ratio (PSNR), utilized in syntactic communication, fail to adequately measure semantic-level distortion. Recently a study have also echoed this sentiment and offering methods to evaluate semantic distortion. The authors in \cite{Blau_2018} contemplate the impact of perception on the quality of restored images amidst lossy compression, thereby extending the classic rate-distortion function. Following this, within \cite{Blau2019RethinkingLC}, they proceed to derive the Rate-Distortion-Perception Function (RDPF), which first comprehensively demonstrated the tradeoff between perception and bit-level distortion. Then the authors in \cite{Theis2021ACT} elucidate the viability of the aforementioned function through the employment of stochastic, variable-length codes. In \cite{Chen2022OnTR}, RDPF has been demonstrated that can be achieved through deterministic codes, except for specific extreme cases. The above enhancement infuses traditional distortion measures with divergence constraints, offering an approach to gauge data satisfaction from a human perspective. It is indicated that minimizing distortion does not necessarily result in perceptually satisfying outputs for human subjects. In other words, a reduction in bit-level distortion does not inherently enhance perceptual quality. Measurement of perceptual quality can be achieved through the divergence between the probability distributions of messages. 

However, the current RDPF has not been validated through real-channel experiments in the integration of semantic communication, and RDPF is a constrained optimization, which is not easy to converge to a global optimal solution. To address the challenges mentioned, this study employs JSCC schemes to construct a semantic communication model, which focus on transmitting semantic information rather than bit-level data. Subsequently, by employing the Information Bottleneck (IB) method\cite{article}, an explicit rate-distortion-bottleneck trade-off is established, transforming the original rate-distortion trade-off from a constrained optimization problem into an unconstrained optimization problem. Additionally, we integrate the perceptual quality of semantic information into the IB approach. Consequently, we develop a rate-distortion-perception-bottleneck (RDPB) model based on the IB framework, and the experimental results based on the image-based semantic communication model have been presented to corroborate our theoretical findings.

\section{System Model And Rate-Distortion-Perception-Bottleneck}

\subsection{System Model}
We investigate a task-oriented communication within a device-edge collaborative inference system. As shown in Fig. \ref{ Semantic Communication model}, we assume the deployment of DNNs on both edge servers and terminal equipment, enabling them to collaboratively execute inferential tasks. The input data $\boldsymbol{x}$ and its corresponding target variable $\boldsymbol{y}$ are considered as separate realizations of a pair of random variables $(X,Y)$.
The encoded feature, received feature, inference result, and restored image are denoted by the random variables $Z$, $\hat{Z}$, $\hat{Y}$ and $\hat{X}$, respectively. The following probabilistic graphical model demonstrates the relationship between the above random variables:
\begin{equation}
\small
\begin{tikzcd}[row sep=tiny]
 & & & & \hat{X}  \\
Y \arrow[r] & X  \arrow[r] & Z \arrow[r] & \hat{Z} \arrow[ur, sloped] \arrow[dr, sloped] & \\
 & & & & \hat{Y}
\end{tikzcd}
\end{equation}
\\which satisfies $p(\boldsymbol{\hat{y}}, \boldsymbol{\hat{x}}, \boldsymbol{\hat{z}}, \boldsymbol{z} | \boldsymbol{x}) = p_\theta(\boldsymbol{\hat{y}}|\boldsymbol{\hat{z}},\boldsymbol{\hat{x}}) p_\eta(\boldsymbol{\hat{x}}|\boldsymbol{\hat{z}},\boldsymbol{\hat{y}}) p_{\text{channel}}\\
(\boldsymbol{\hat{z}}|\boldsymbol{z}) p_{\phi}(\boldsymbol{z}|\boldsymbol{x})$
, parameterized by $\theta$, $\eta$ and $\phi$ of the DNNs to be further elaborated in the following sections.

The device’s network delineates a conditional distribution $p_\phi(\boldsymbol{z} | \boldsymbol{x})$ with the parameter $\phi$,  which encompasses both a feature extractor and a JSCC encoder. Initially, the extractor discerns the task-specific feature input raw data $\boldsymbol{x}$, subsequently maping these feature values into the channel input symbol $\boldsymbol{z}$ via the JSCC encoder. Given that both the extractor and encoder are parameterized by DNNs, it is feasible to conduct joint training of these two modules in an end-to-end manner. And considering the need for digital signal transmission in channel communication, we have implemented uniform quantization on $z$ in this scenario, meanwhile we employ the output dimension $dim$ of the encoder and the quantization level $L$ to control rete, thereby obtaining $R = dim *\log_{2}(L)$. Subsequently, the encoded feature $z$ is conveyed to the receiver via the noisy wireless communication channel, where the receiver acquires the noise-distorted feature $\hat{\boldsymbol{z}}$. For the sake of simplicity, we assumes a additive white Gaussian noise (AWGN) channel between the source and the receiver, represented as an untrainable layer with a transfer function denoted by $\boldsymbol{\hat{z}} = \boldsymbol{z} + \epsilon$. The channel noise $\epsilon$ is assumed to be Gaussian with a zero mean and a variance of $\sigma^2$, which can be denoted as $\epsilon \sim N(0 , \epsilon^2 \boldsymbol{I})$. The final features $\boldsymbol{z}$ are respectively reconstructed into the original image $\boldsymbol{\hat{x}}$ via the JSCC decoder with the distribution $p_\eta(\boldsymbol{\hat{x}} | \boldsymbol{\hat{z}},\boldsymbol{y})$ parameterized by $\eta$ and inferred outputs $\boldsymbol{\hat{y}}$ through the inference network with the distribution $p_\theta(\boldsymbol{\hat{y}} | \boldsymbol{\hat{z}},\boldsymbol{x})$ parameterized by $\theta$. Therefore, in our model, there are two types of distortion: one stemming from the image reconstruction network, and the other from the image inference network.

\subsection{Rate-Distortion-Perception-Bottleneck}
The overhead of communication is related to the feature extraction vector $Z$ with non-zero dimensions. In general, transmitting higher-dimensional feature vectors $Z_n$ leads to better transmission accuracy. In accordance with this paper’s investigation\cite{Blau2019RethinkingLC}, we have incorporated them RDP into our methodology. Similar with \cite{shao2021learning}, we employ the IB approach to transform the constrained RDP optimization function into an unconstrained RDPB optimization function, with the objective of minimizing the following target function:
\begin{equation}
\begin{aligned}
\label{rdp}
\mathcal{L}_{RDPB} = &-I(\hat{Z},Y)+\beta I(\hat{Z},X)+\lambda
\mathbb{E}[\Delta(X, \hat{X})] \\
&\quad\quad\quad\quad\text{s.t.} \quad d(X,\hat{X}) \leq P
\end{aligned}
\end{equation}
At a constant rate, the encoder-decoder is trained to minimize a loss function that consists of a weighted blend of the anticipated distortion and the quality of perception. The perceptual quality of an image $\hat{x}$ is deqfined by its resemblance to a natural image, unrelated to its similarity to any reference \cite{Blau_2018}. In compression domains, the metric for perceptual quality can manifest as a discrepancy between distributions, such as the TV distance, KL divergence, Wasserstein distance, etc. In this paper, we choose the KL divergence $ D_{KL}(X,\hat{X})$ to measure the distance between $X$ and $\hat{X}$. 
\begin{equation}
D_{KL}(p(\boldsymbol{x}) | q(\boldsymbol{\hat{x}})) = \int p(\boldsymbol{x}) \log \frac{p(\boldsymbol{x})}{q(\boldsymbol{\hat{x}})} d\boldsymbol{x}
\end{equation}
Therefore, we obtained the following objective function:
\allowdisplaybreaks
\begin{equation}
\begin{aligned}
 \label{RDPB}
\mathcal{L}_{RDPB}(\theta) =& \mathbf{E}_{p(\boldsymbol{x},\boldsymbol{y})}\big\{ \mathbf{E}_{p_{\phi}(\boldsymbol{\hat{z}} | \boldsymbol{x})}[-\log p(\boldsymbol{y} | \boldsymbol{\hat{z}})] \\[8pt]
&+ \beta D_{K L}\left(p_{\boldsymbol{\phi}}(\boldsymbol{\hat{z}} | \boldsymbol{x}) \| p(\boldsymbol{\hat{z}})\right)\big\} \\[6pt]
&+ \lambda\mathbb{E}[\Delta(X, \hat{X})]+\mu D_{KL}\left(p_{\boldsymbol{X}}, p_{\boldsymbol{\hat{X}}}\right)
\end{aligned}
\end{equation}
where $p(\boldsymbol{\hat{z}})$ is the true marginal distribution of received feature and $p(\boldsymbol{\hat{z}}|\boldsymbol{x})$ is the posterior distribution given $(\boldsymbol{x},\boldsymbol{y})$, and the proof please refer to the Appendix of \ref{Appendix:rdpb}. The objective function comprises a weighted combination of four terms, where $\beta, \lambda, \mu > 0 $ govern the tradeoff. Specifically, the quantity $I(\hat{Z},X)$ can be understood as the information being transferred from $X$ to $Z$ and stored by $\hat{Z}$ after being disturbed by noise, which can be measured by rate (can be obtained by multiplying the digitalizing bit by the length of the feature vector). As $H(Y)$ is a constant related to the input data, it can be omitted in the last line to optimize the loss function. Furthermore, we can deduce that minimizing either $-I(\hat{Z},Y)$ or $H(Y|\hat{Z})$ is equivalent outcomes, which can be comprehended as the distortion introduced by the noise-corrupted feature vector $\hat{Z}$ in inferring $\hat{Y}$. $d_{TV}(X,\hat{X})$ represents the perceptual discrepancy between the quality of the reconstructed image $\hat{X}$ and the original image $X$. Although the image reconstruction section and the image classification section are formed by two branches of networks, minimizing $d_{TV}(X,\hat{X})$ can enhance the representational capacity of the feature vector $\hat{Z}$. Therefore, we leveraging the IB principle formalizes a RDP tradeoff in semantic communication systems, culminating in the development of RDPB. In comparison to the RDPF, the use of the IB framework preserves task-relevant information, resulting in $I(\hat{Z},X)$ significantly smaller than $H(X)$, reducing communication overhead, and finally leading to an unrestricted loss function.
\begin{figure*}[t]
\centering
\subfloat[]{{
  \centering
  \includegraphics[width=0.48\textwidth]{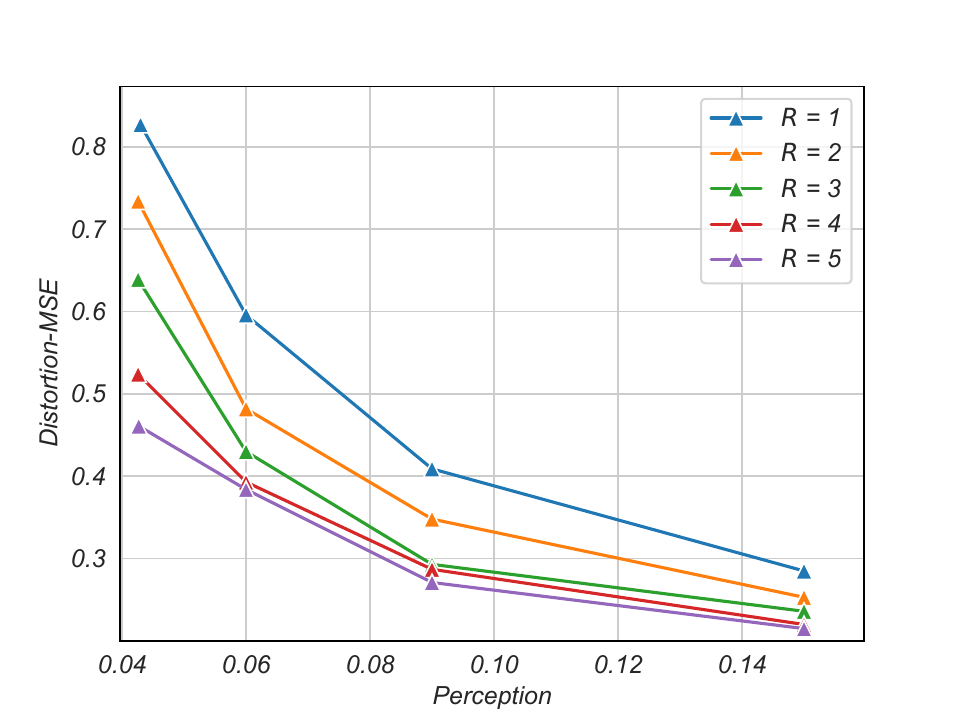}
  \label{RDPB(a)}
}}
\subfloat[]{{
\centering
\includegraphics[width=0.48\textwidth]{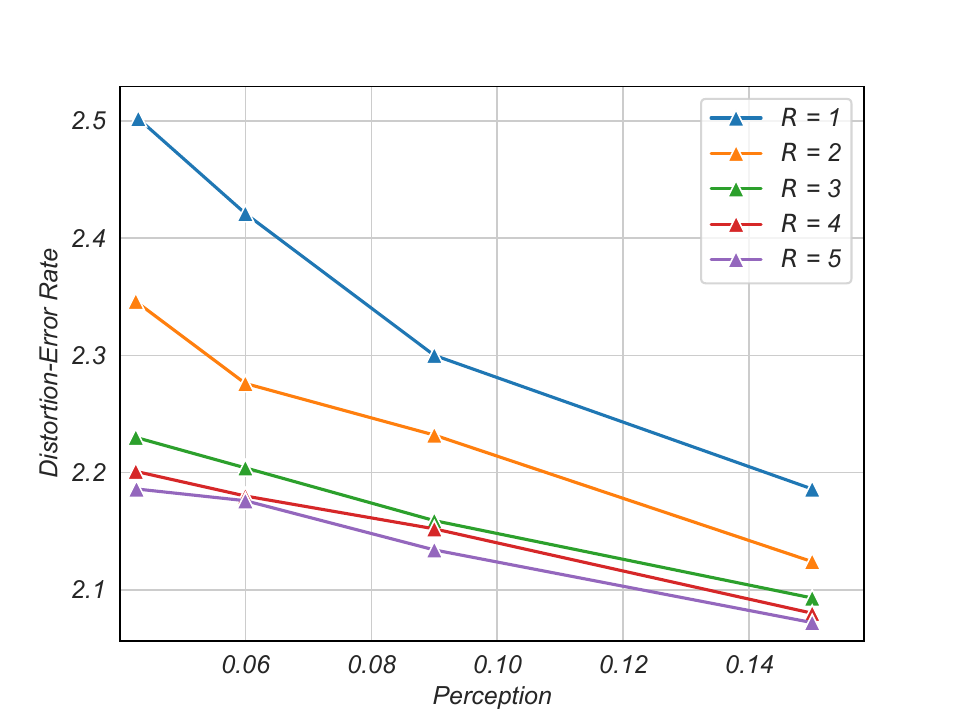}
\label{RDPB(b)}
}}
\\
\subfloat[]{{
\centering
\includegraphics[width=0.48\textwidth]{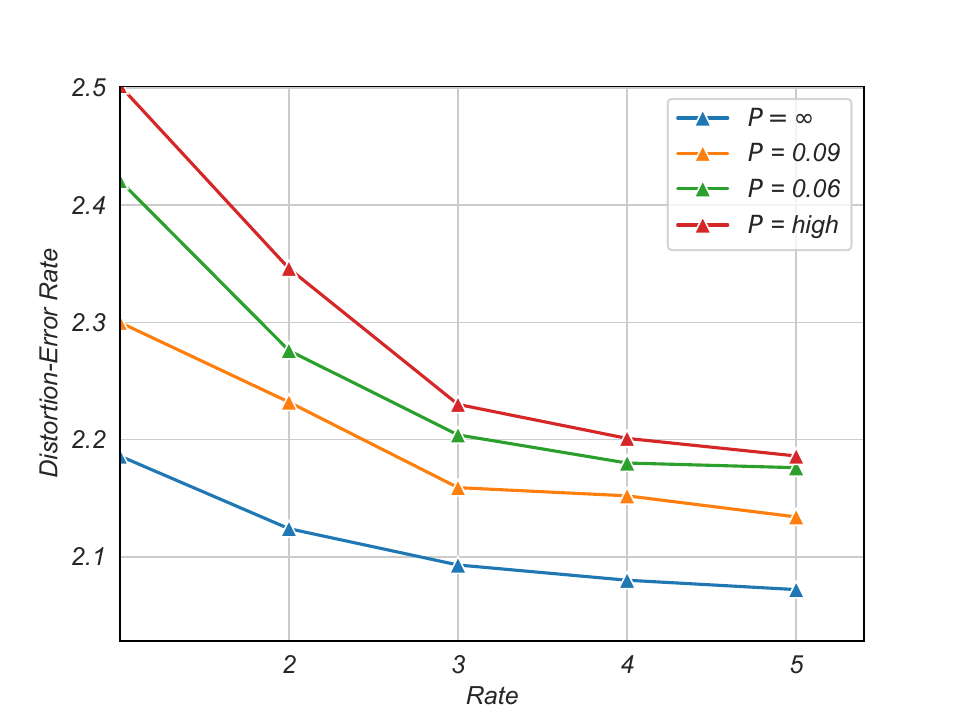}
\label{RDPB(c)}
}}
\subfloat[]{{
\centering
\includegraphics[width=0.48\textwidth]{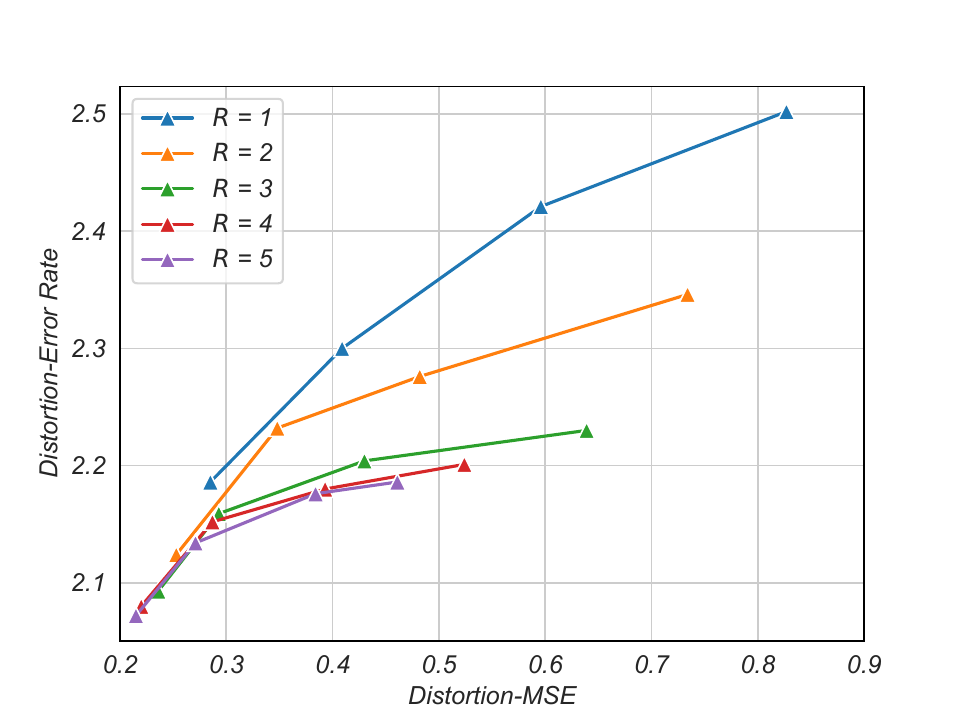}
\label{RDPB(d)}
}}

\caption{The rate-distortion-perception-bottleneck function of MNIST images.}
\label{RDPB_figure}

\end{figure*}

In the preceding text, with a given distribution $p_\phi(\boldsymbol{\hat{z}},\boldsymbol{x})$, the joint data distribution $p(\boldsymbol{x},\boldsymbol{y})$, the distributions $p(\boldsymbol{\hat{z}})$ and $p(\boldsymbol{y}|\boldsymbol{\hat{z}})$ are completely determined by the underlying Markov chain $Y\xleftarrow{}X\xrightarrow{}\hat{Z}$. However, these distributions become computationally challenging due to the presence of high-dimensional integrals:
\begin{equation}
     p(\bm{\hat{z}}_{k}) =\sum_{x \in X} p(\bm{x}) p_{\bm{\theta}}(\bm{\hat{z}} | \bm{x}) 
\end{equation}

\begin{equation}
     p(\bm{y} | \bm{\hat{z}}) = \sum_{x \in X} \frac{ p(\bm{x}, \bm{y}) p_{\bm{\theta}}(\bm{\hat{z}} | \bm{x})  }{p(\bm{\hat{z}})} 
\end{equation}

\begin{figure*}[t]
\centering
\subfloat[Accuracy  $= 97.65\%$ , $R = 2$, $P = best$]{
\centering
\includegraphics[width=0.45\textwidth]{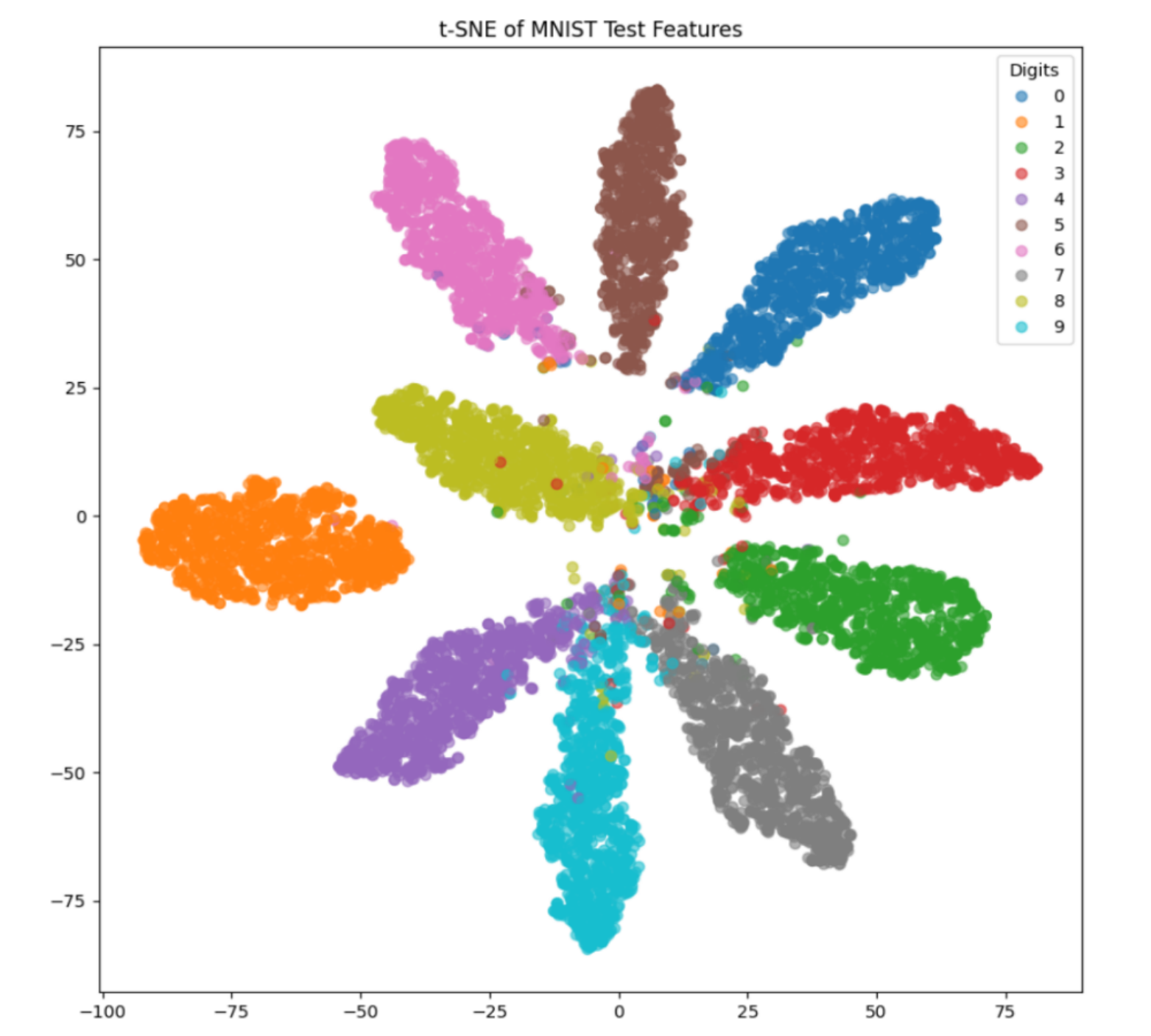}
}
\subfloat[Accuracy  $= 97.88\%$ , $R = 2$, $P = \infty$ ]{
\centering
\includegraphics[width=0.45\textwidth]{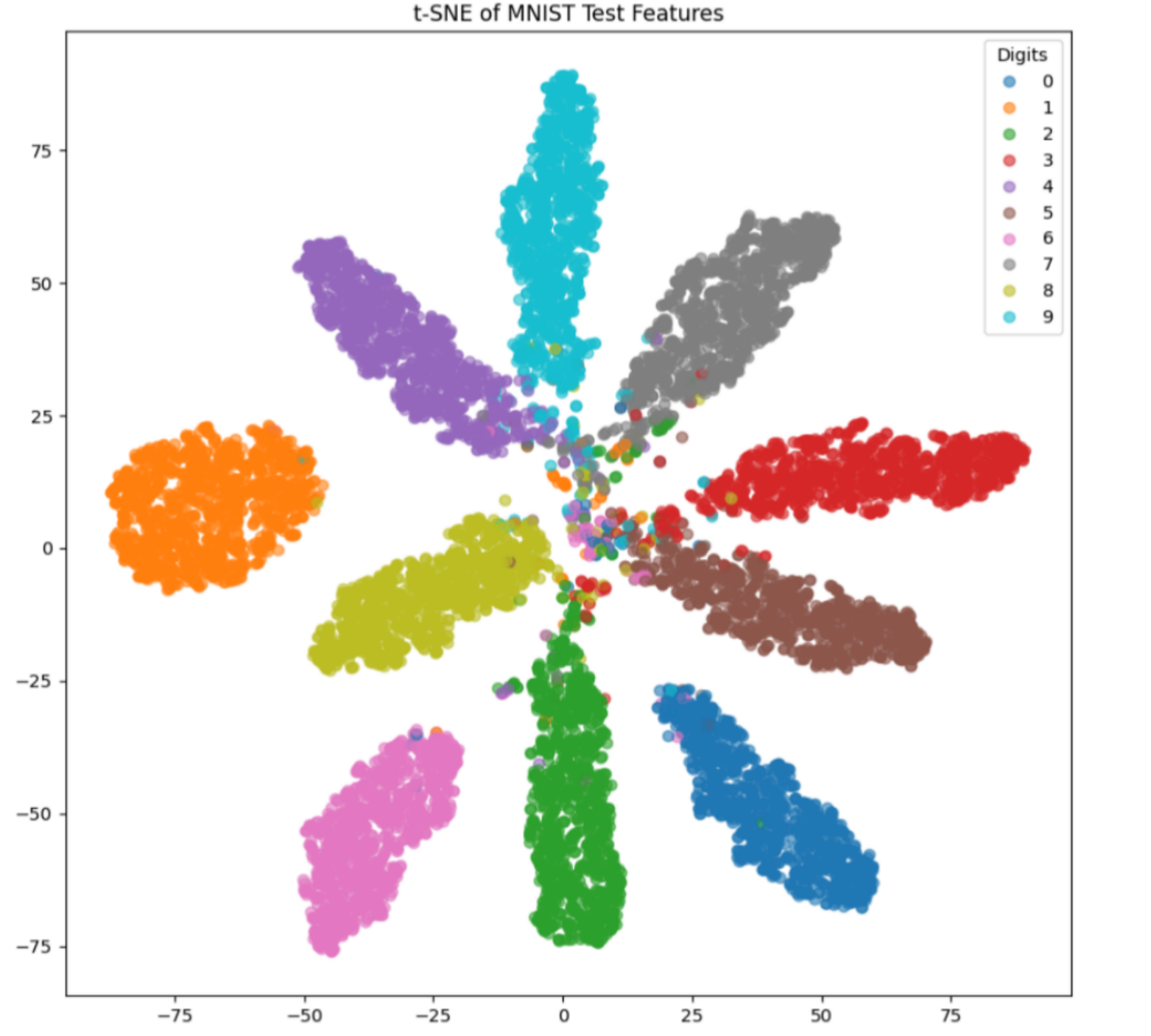}
}
\caption{two-dimensional $t$-SNE embedding of the received feature in the MNIST classification task.}
\label{t-sne}
\end{figure*}

In order to handle the high dimensionality of the distributions $p(\bm{z})$ and $p(\bm{y|\hat{z}})$, we choose to utilize variational inference tools to approximate these challenging distributions and 
derive the rate distortion perception variational bottleneck (RDPVB) function. The concept of variational approximation involves positing a family of distributions and identifying a member within that family that closely resembles the target distribution \cite{Blei_2017}. In particular, we introduce $q(\bm{z})$ and $q(\bm{y|\bm{\hat{z}}})$ as variational distributions to provide approximations for $p(\bm{z})$ and $p(\bm{y|\bm{\hat{z}}})$, respectively. By using these approximations, the objective function can be transformed as follows:
\begin{equation}
\label{eq:rdpb-vib}
\begin{aligned}
    \mathcal{L}_{RDPB}(\theta,\phi) =& \mathbf{E}_{p(\boldsymbol{x},\boldsymbol{y})}\big\{ \mathbf{E}_{p_{\phi}(\boldsymbol{\hat{z}} | \boldsymbol{x})}[-\log q_\theta(\boldsymbol{y} | \boldsymbol{\hat{z}})] \\[8pt]
        &+ \beta D_{K L}\left(p_{\phi}(\boldsymbol{\hat{z}} | \boldsymbol{x}) \| q(\boldsymbol{\hat{z}})\right)\big\} \\[6pt]
        &+ \lambda\mathbb{E}[\Delta(X, \hat{X})]+\mu D_{KL}\left(p_{\boldsymbol{x}}, p_{\boldsymbol{\hat{x}}}\right)
\end{aligned}
\end{equation}
A comprehensive derivation is presented in Appendix \ref{Appendix:rdpb_vib}. To optimize the negative log-likelihood term in (\ref{eq:rdpb-vib}), we utilize the reparameterization trick \cite{Kingma2013AutoEncodingVB} and Monte Carlo sampling to obtain an unbiased estimate of the gradient, thereby enhancing our computational results. This allows us to utilize stochastic gradient descent to optimize the objective. Specifically, given a small batch of data ${(\bm{x_i},\bm{y_i})}_{i=1}^M$ , we have the following unbiased estimation:
\begin{equation}
\label{eq:rdpb-vib-simeq}
\begin{aligned}
\mathcal{L}_{RDPVB}(\theta,\phi) \simeq & \frac{1}{M} \sum_{m=1}^{M}
\big\{ \left[-\log q_{\theta}\left(\boldsymbol{y}_{m} \mid \boldsymbol{\hat{z}}_{m, l}\right)\right] \\[6pt]
&+\beta D_{K L}\left(p_{\phi}\left(\boldsymbol{\hat{z}} \mid \boldsymbol{x}_{m}\right) \| q(\boldsymbol{\hat{z}})\right)\\[6pt]
&+ \lambda\mathbb{E}[\Delta(X, \hat{X})]+\mu D_{KL}\left(p_{\boldsymbol{x}}, p_{\boldsymbol{\hat{x}}}\right) \big\}\\[3pt]
\end{aligned}
\end{equation}

\section{Experimental Illustration}
\subsection{Experiments Setup}
We now turn to demonstrate the RDPB in semantic communication on a toy MNIST example\cite{article}. The MNIST dataset is composed of a series of grayscale images each representing a handwritten digit from “0” to “9”. The dataset consists of two main components: a training set and a test set. Specifically, the training set contains 50,000 examples, while the test set comprises 10,000 examples. We evaluate the distortion and perceptual quality of the model under varying desired configurations on the MNIST dataset, thereby obtaining the experimental RDPB curve.
In image classification networks, we utilize classification accuracy as a measure of inference performance, which corresponds to "distortion". While for image reconstruction networks, we use MSE as the distortion of this network. And we employ the quantization level $L$ and the output dimension $dim$ of the encoder to control rete $R$. Then we attain different balances between distortion and perceptual quality by controlling the hyperparameter $\lambda$, $\beta$, $\mu$.

To control the complexity of the model, we begin by training the image classification network. While the image classification network has been trained to a satisfactory level , then we introduce the image reconstruction network into the training process.

Given that we employ the KL divergence to quantify the semantic perception, and that KL divergence utilizes logarithmic computations, its calculated value reaches infinity within the traditional Shannon communication framework. To facilitate graphical representation and comparison with other experimental outcomes, we assign Shannon’s perception a relatively substantial value of 0.15. Meanwhile, in an effort to render the outcomes of our experiment both clearer and more exhaustive, we select two specific values, namely $perception = 0.06$ and $perception = 0.09$, to represent the medium cases of perception. Should the perception fall below these thresholds during the training phase, we then proceed to eliminate the KL divergence from the loss function.

\subsection{Experiments Results}
In Fig \ref{RDPB(a)}, \ref{RDPB(b)}, the curves illustrate the trade-off between distortion and perceived quality, irrespective of whether the distortion pertains to the Distortion-Error Rate in image classification tasks or the Distortion-MSE in image reconstruction endeavors. It is observable that this compromise intensifies at lower bit rates. 

Fig \ref{RDPB(c)} depicts the enhancement of perceived quality through rate elevation amidst continuous distortion. Fig \ref{RDPB(d)} elucidates the interrelation between the Error Rate in image classification tasks and the MSE in image reconstruction tasks. And drawing from the insights garnered through preliminary experiments, it is observed that integrating the task of image restoration with the incorporation of MSE into the loss function can result in a diminution of error rates.

Moreover, we additionally illustrate the noisy feature vector $\hat{z}$ within the MNIST classification endeavors in Figure \ref{t-sne} via a two-dimensional $t$-distributed stochastic neighbor embedding ($t$-SNE) \cite{JMLR:v9:vandermaaten08a}. Despite the accuracy on the left being inferior to that on the right, the enhanced semantic awareness trained in the left-hand illustration permits the inference network to extract information of superior quality from the feature vector $\hat{z}$, thereby resulting in a more favorable distribution of data points compared to the right-hand illustration.

\begin{figure*}[t]
\normalsize
\begin{equation}
\label{appendix_rdpb}
\begin{aligned}
\mathcal{L}_{RDPB}(\theta) =& \underbrace{-I(\hat{Z}, Y) + \lambda
\mathbb{E}[\Delta(X, \hat{X})]}_{\text {Distortion }}+ \beta  
\underbrace{I(\hat{Z}, X)}_{\text {Rate }} + \mu  \underbrace{d(X , \hat{X})}_{\text {Perception }}\\
=& \mathbf{E}_{p(\boldsymbol{x},\boldsymbol{y})}\big\{ \mathbf{E}_{p_{\phi}(\boldsymbol{\hat{z}} | \boldsymbol{x})}[-\log p(\boldsymbol{y} | \boldsymbol{\hat{z}})] + \beta D_{K L}\left(p_{\phi}(\boldsymbol{\hat{z}} | \boldsymbol{x}) \| p(\boldsymbol{\hat{z}})\right)\big\} - H(Y) + \lambda \mathbb{E}[\Delta(X, \hat{X})]+ \mu D_{KL}\left(p_{\boldsymbol{x}}, p_{\boldsymbol{\hat{x}}}\right) \\[6pt] 
=& \mathbf{E}_{p(\boldsymbol{x},\boldsymbol{y})}\big\{ \mathbf{E}_{p_{\boldsymbol{\phi}}(\boldsymbol{\hat{z}} | \boldsymbol{x})}[-\log p(\boldsymbol{y} | \boldsymbol{\hat{z}})] + \beta D_{K L}\left(p_{\boldsymbol{\phi}}(\boldsymbol{\hat{z}} | \boldsymbol{x}) \| p(\boldsymbol{\hat{z}})\right)\big\} + \lambda\mathbb{E}[\Delta(X, \hat{X})]+\mu D_{KL}\left(p_{\boldsymbol{x}}, p_{\boldsymbol{\hat{x}}}\right)\\[8pt] 
\end{aligned}
\end{equation}
\hrulefill
\end{figure*}

\begin{figure*}[t]
\normalsize
\begin{equation}
\label{equ_appendix_rdpvb}
\begin{aligned}
\mathcal{L}_\mathrm{RDPB}(\phi) = &-\sum_{y \in Y} \sum_{\hat{z} \in \hat{Z}} p(y|\hat{z}) p(\hat{z}) \log \frac{p(y|\hat{z})}{p(y)} + \beta \sum_{\boldsymbol{x} \in X} \sum_{\boldsymbol{\hat{z}} \in \hat{Z}} p_{\phi}(\boldsymbol{\hat{z}}|\boldsymbol{x}) p(\boldsymbol{x}) \log \frac{p_{\phi}(\boldsymbol{\hat{z}}|\boldsymbol{x})}{p(\boldsymbol{\hat{z}})} + \lambda\mathbb{E}[\Delta(X, \hat{X})]+\mu D_{KL}\left(p_{\boldsymbol{x}}, p_{\boldsymbol{\hat{x}}}\right)\\
= &-\sum_{\boldsymbol{y} \in Y} \sum_{\boldsymbol{\hat{z}} \in \hat{Z}} p(\boldsymbol{y}|\boldsymbol{\hat{z}}) p(\boldsymbol{\hat{z}}) \log q_\theta(\boldsymbol{y}|\boldsymbol{\hat{z}}) + \beta \sum_{\boldsymbol{x} \in X} \sum_{\boldsymbol{\hat{z}} \in \boldsymbol{\hat{Z}}} p_{\phi}(\boldsymbol{\hat{z}|x)} p(\boldsymbol{x}) \log \frac{p_{\phi}(\boldsymbol{\hat{z}}|\boldsymbol{x})}{q(\boldsymbol{\hat{z}})} + \lambda\mathbb{E}[\Delta(X, \hat{X})]+\mu D_{KL}\left(p_{\boldsymbol{x}}, p_{\boldsymbol{\hat{x}}}\right)\\
= &-D_{KL}(p(\boldsymbol{y}|\boldsymbol{\hat{z}}) || q_\theta(\boldsymbol{y}|\boldsymbol{\hat{z}})) - \beta D_{KL}(p(\boldsymbol{\hat{z}}) || q(\boldsymbol{\hat{z}})) + \lambda\mathbb{E}[\Delta(X, \hat{X})]+\mu D_{KL}\left(p_{\boldsymbol{x}}, p_{\boldsymbol{\hat{x}}}\right)
\end{aligned}
\end{equation}
\hrulefill
\end{figure*}

\section{Conclusion}

In this work, we employ JSCC to construct a semantic communication model. By utilizing the IB framework, we transform the conventional constrained RDP tradeoff into an unconstrained RDPB tradeoff. This function simplifies the optimization process, facilitating convergence to the global optimum more readily and obtains the tradeoff among the data rate, symbol distortion, and semantic perception in semantic communication. Extensive simulations on the MNIST image classification dataset substantiate the effectiveness of our proposed method.

\appendices
\section{Motivated function}
\label{Appendix:rdpb}

Revisit the RDP objective in (\ref{rdp})  $\mathcal{L}_{RDPB} = -I(\hat{Z},Y)+\beta I(\hat{Z},X)+\lambda
\mathbb{E}[\Delta(X, \hat{X})] \quad\text{s.t.}  d(X,\hat{X}) \leq P$. Through the derivation shown in (\ref{appendix_rdpb}), we can arrive at a specific formulaic expression.

\section{Derivation of the Variational Upper Bound}
\label{Appendix:rdpb_vib}

Recall that the RDPB objective in (\ref{RDPB}) has the form $\mathcal{L}_\mathrm{RDPB}(\gamma) = -I(\hat{Z}, Y) + \lambda
\mathbb{E}\Delta(X, \hat{X})+ \beta  I(\hat{Z}, X)+ \mu  d(X , \hat{X})$. Writing it out in full with the conditional distribution $p_{\bm{\theta}}(\bm{z}|\bm{x})$ and the variational distributions $p_{\bm{\varphi}}(\bm{y}|\bm{z})$, $r(\bm{z})$, the derivation is shown in (\ref{equ_appendix_rdpvb}). $\mathcal{L}_\mathrm{RDPVB}(\phi)$ in the formulation is the RDPVB objective function in (\ref{eq:rdpb-vib-simeq}). Given the non-negativity of the KL-divergence, the RDPVB objective function serves as an upper limit of the RDPB objective function.

\bibliographystyle{IEEEtran}
\begin{spacing}{1.1} 
    \bibliography{ref}   
  \end{spacing}

\end{document}